\documentclass[%
 aip,
 amsmath,amssymb,
 reprint,%
]{revtex4-1}

\usepackage{graphicx}
\usepackage{dcolumn}
\usepackage{bm}

\usepackage[utf8]{inputenc}
\usepackage[T1]{fontenc}
\usepackage{mathptmx}
\usepackage[labelformat=empty,position=top]{subcaption}
\usepackage[font=footnotesize,labelfont=small,singlelinecheck=off,justification=raggedright]{caption}
\begin{document}

\preprint{AIP/123-QED}

\title{Effects of Proton Irradiation on Hole Carrier Transport in Hydrogen-Terminated Diamond Surfaces}

\author{R. Peterson}
\affiliation{Department of Electrical Engineering, Stanford University, California 94305, USA}
\author{M. Malakoutian}
\affiliation{Department of Electrical and Computer Engineering, University of California, Davis, California 95616,USA}
\author{Y. Wang}
\affiliation{Material Science and Technology Division, Los Alamos National Laboratory, Los Alamos, New Mexico 87545,USA}
\author{S. Chowdhury$^{1,}$}
\affiliation{Department of Electrical and Computer Engineering, University of California, Davis, California 95616,USA}
\author{D. Senesky}\thanks{Electronic mail: dsenesky@stanford.edu}
\affiliation{Department of Aeronautics and Astronautics, Stanford University, California 94305,USA}

\begin{abstract}
In this letter, we report the effects of irradiating hydrogen-terminated diamond surfaces with protons. We study the electrical behavior of the two-dimensional hole gas that forms on the diamond surface as a result of hydrogen-termination. Four diamond samples, two of which were passivated with Al$_2$O$_3$ and the other two unpassivated, were irradiated with 2 MeV protons at two fluences, 0.5$\times$10$^{14}$ and 1.0$\times$10$^{14}$~cm$^{-2}$. The hole conductivity degraded at a higher proton fluence, which is attributed to the reduction of the mobility caused by enhanced hole scattering. Both passivated and unpassivated samples exhibited a reduction in the mobility, which can be explained by charging effects of the Al$_2$O$_3$ (for the passivated samples) and degradation of the hydrogen-terminated surface (for the unpassivated samples). To our knowledge, this is the first reported data on 2 MeV proton tolerance of hydrogen-terminated diamond surfaces. 

\end{abstract}

\maketitle

Diamond-based electronics have been investigated for many years, showing promising performance for high-power and high-frequency applications, owing to its figures of merit (FOM). Such FOM values are high due to diamond's wide band gap, high breakdown voltage, high thermal conductivity, and high saturation velocity. Moreover, diamond's relatively small lattice constant makes it incredibly hardened, thus exhibiting high resistance to particle radiation. The energy required to permanently displace a carbon atom from the diamond's lattice is considerably higher than other semiconductors \cite{pearton2015ionizing,ionascut2002radiation,nava2008silicon}. This makes diamond-based devices an attractive candidate for radiation-rich environments. However, most of the work of irradiated diamond has focused on radiation-monitoring applications, which usually consists of two electrodes for biasing and charge collection~\cite{proton_irrad1,proton_irrad2,proton_irrad3,proton_irrad4,proton_irrad5,proton_irrad6}. This includes both charged ion irradiation and neutron irradiation~\cite{neutron1,neutron2,neutron3,verona2016_neutrons}. Additionally, one paper studied the effects of H:diamond FETs under neutron irradiation, which concluded that the operation post-irradiation was robust and degradation effects were minuscule or undetectable~\cite{verona2016_neutrons}. 

Unlike neutrons, however, penetrating charged particles interact strongly with the electron subsystem of a lattice via ionization, and such particles are highly abundant in space environments. For hydrogen-terminated diamond, several questions arise in the context of charged particle irradiation. Although the threshold atomic displacement energy is high for diamond, the ionization that occurs as a charged particle penetrates the lattice may cause damage and thus impact the electrical performance.

Using SRIM, we can show that the energy deposition rate via ionization (i.e., electronic stopping power) for a 2~MeV proton penetrating a diamond lattice is 50~eV/nm, or about $\sim$18~eV for every lattice constant of diamond \cite{ziegler2010srim}. This energy deposition rate occurs locally along the proton trajectory, which produces a cylindrical ionization volume. However localized that these effects may be, it is important to determine if the integral fluence is sufficient to modify the electrical properties. Thus, in this work, we study this by proton-irradiating four H:diamond samples at two different fluences.

To study the oxide TID effects, two of the irradiated samples were passivated Al$_2$O$_3$/H:diamond structures with fabricated metal-insulator transistors and capacitors (MISFETs and MISCAPs), as well as Hall-effect devices. It is worth noting, however, that any degradation intrinsic to the H:diamond structure may be difficult to decouple from the electrical impact induced by the degraded oxide. This is particularly the case because the C-H bond uniformity is essential to the conduction of holes \cite{Peterson2020}. For this reason, the other two irradiated samples remained unpassivated. For these samples, Schottky-gated transistors (MESFETs) and Hall-effect devices were used to probe the irradiation effects. This will give us a more holistic understanding of the degradation modes of all H:diamond devices. 

\begin{figure}[t!]\centering
\includegraphics[scale=0.45,trim={0cm 0 0cm 0},clip]{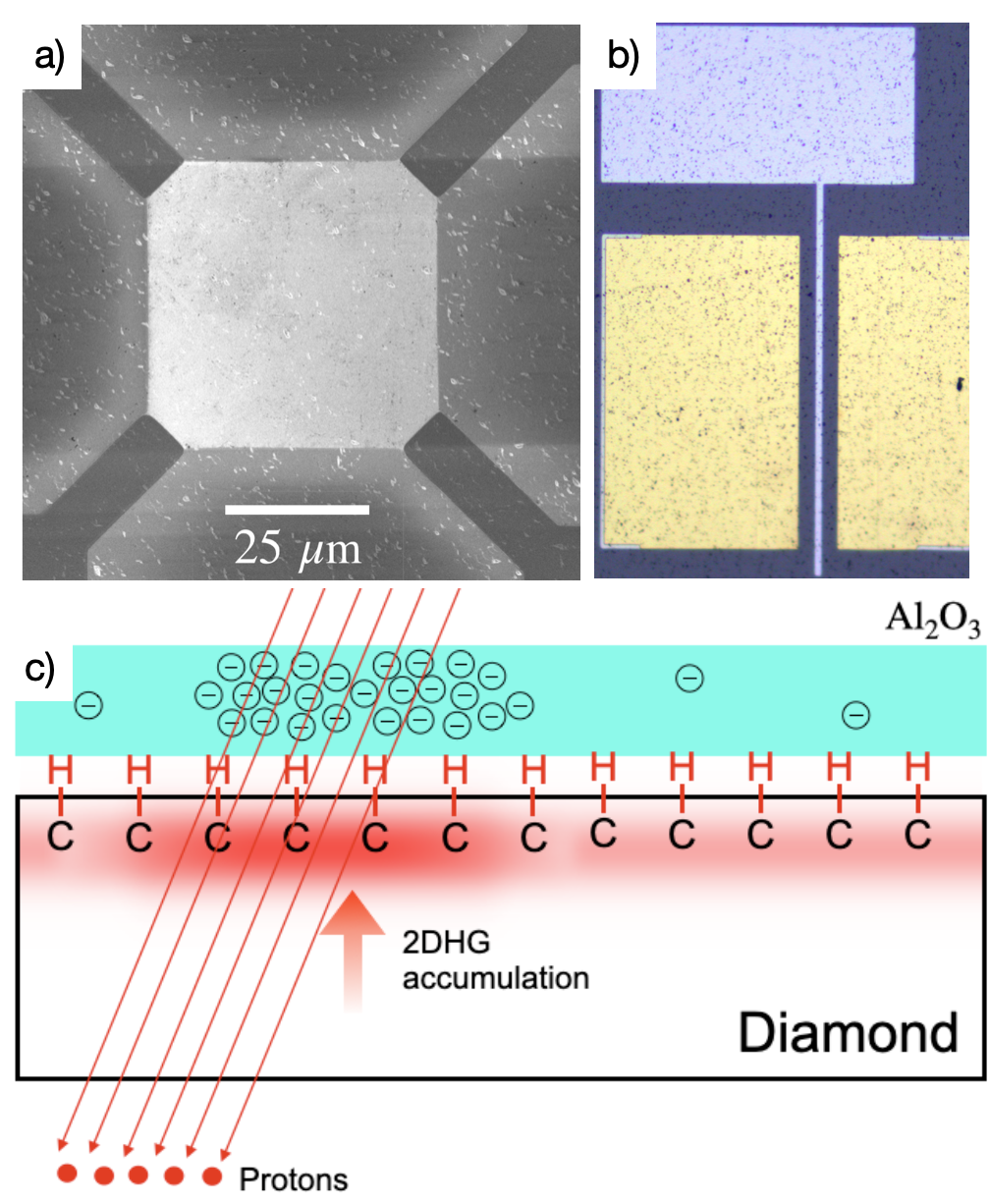}
\caption{a) Scanning electron microscope image of a Hall-effect H:diamond device, b) Optical microscope image of an H:diamond FET, and c) Schematic demonstrating the accumulation of holes due to the negative charge build-up in the Al$_2$O$_3$.}
\label{irrad_al2o3_schematic}
\end{figure}

Two H:diamond samples, which we denote here as AlOx:D1 and AlOx:D2, were passivated simultaneously with 25~nm of Al$_2$O$_3$ via ALD at a 250$^{\circ}$C chamber pressure. This was a thermal deposition process that used trimethylaluminum and H$_2$O as precursors. Prior to the oxide deposition, the two samples were annealed in the ALD chamber for 30 minutes in order to desorb the atmospheric molecules. This ensured that the Al$_2$O$_3$ film was the sole source of acceptors. Once passivated, metal gates (50/20~nm of Al/Pt) were patterned and evaporated. Images of these devices are shown in Fig.~\ref{irrad_al2o3_schematic}(a) and (b).

The proton irradiation was conducted in the Tandem Accelerator at the Ion Beam Materials Laboratory at Los Alamos National Laboratory. The ion fluence measurement was done through a home-made four corner Faraday-cup assembly. The integral proton fluence was 0.5 and 1~$\times 10^{14}$~cm$^{-2}$ for the AlOx:D1 and AlOx:D2 samples, respectively. Using the Al$_2$O$_3$ mass density and the electronic stopping power of 2~MeV protons, this dose converts to 0.95 and 1.9~Mrad, respectively.

Prior to the proton-irradiation, the MISFETs and MISCAPs on the samples were characterized using a B1500 semiconductor parameter analyzer. This included current-voltage and capacitance-voltage measurements. Moreover, the Hall-effect devices were used to extract the hole sheet density and hole mobility. The same measurements were performed after the irradiation exposure. The analysis that follows is on the observed irradiation effects based on these measurements. 

\begin{figure}\centering
\includegraphics[scale=0.48,trim={0cm 0cm 0.5cm 0cm},clip]{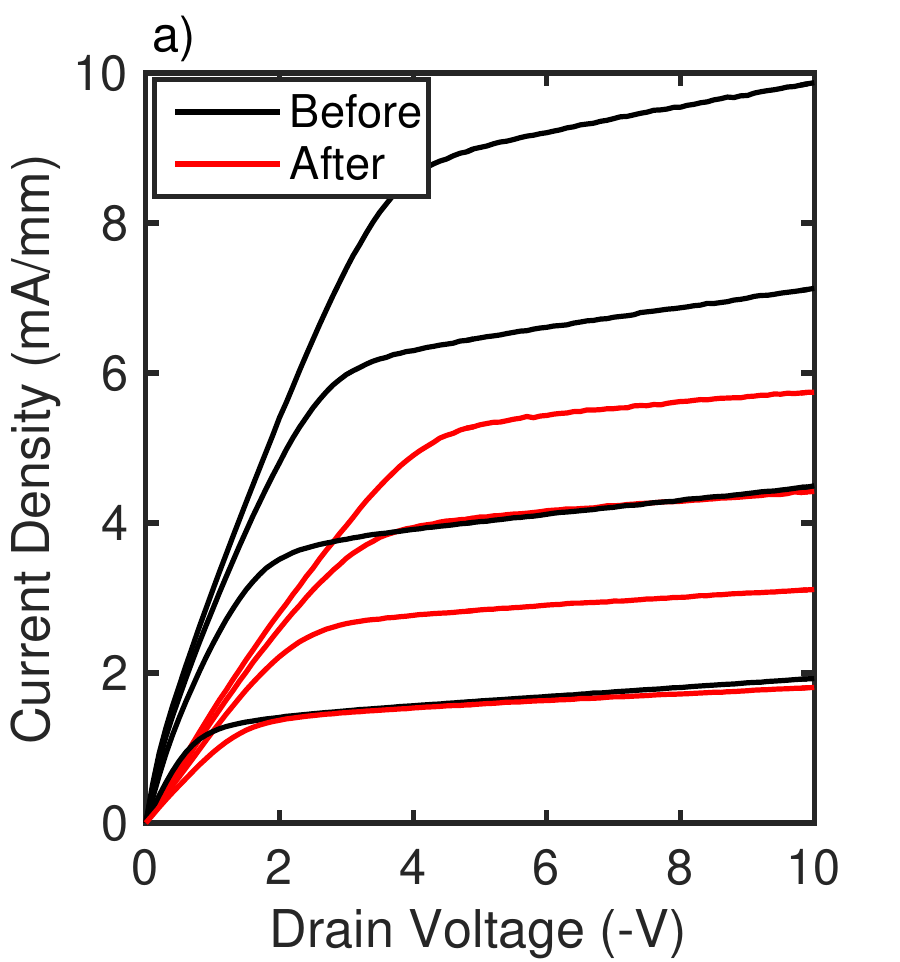}
\includegraphics[scale=0.48,trim={.8cm 0 0cm 0cm},clip]{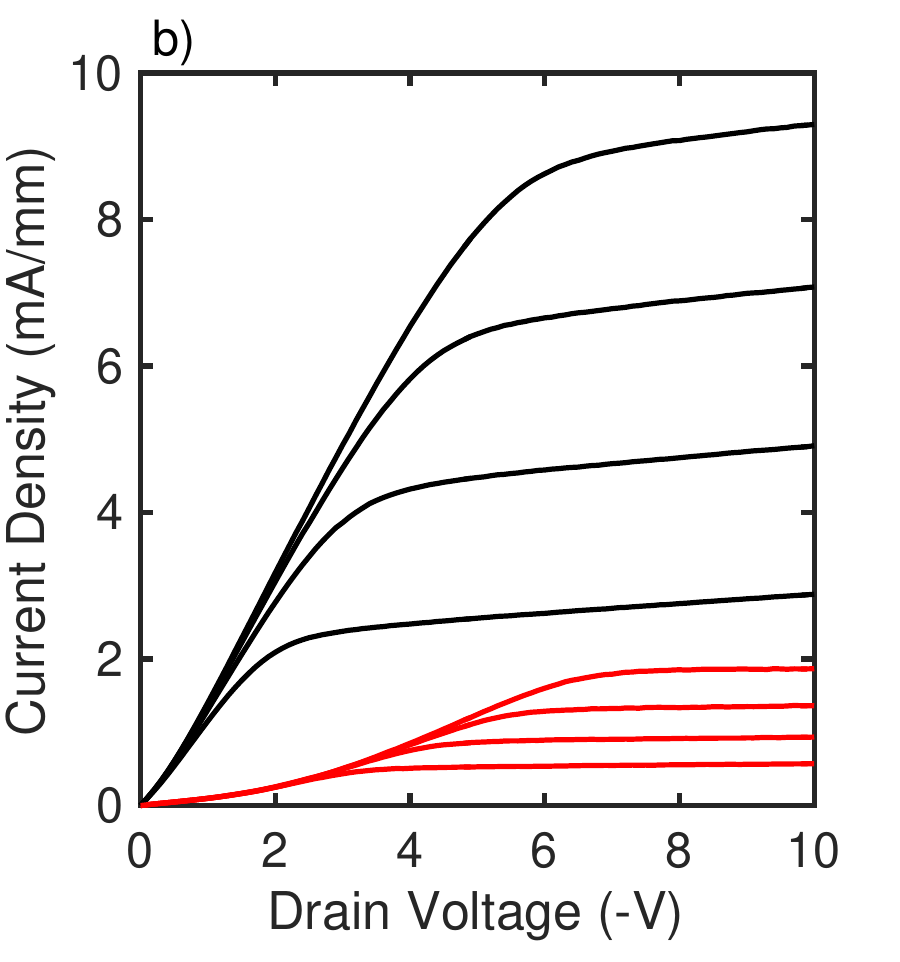}
\caption{ID-VD measurements on the MISFETs before and after irradiation of (a) AlOx:D1, exposed to 0.5~$\times 10^{14}$~protons/cm$^{2}$ and (b) AlOx:D2, exposed to 1~$\times 10^{14}$~protons/cm$^{2}$. These drain currents were measured for gate voltages ranging from -0.5 to -3.5~V, in increments of -1~V.}
\label{SC_al2o3_IDVD_prepostIrrad}
\end{figure}

\begin{figure}\centering
\includegraphics[scale=0.48,trim={0cm 0 0cm 0cm},clip]{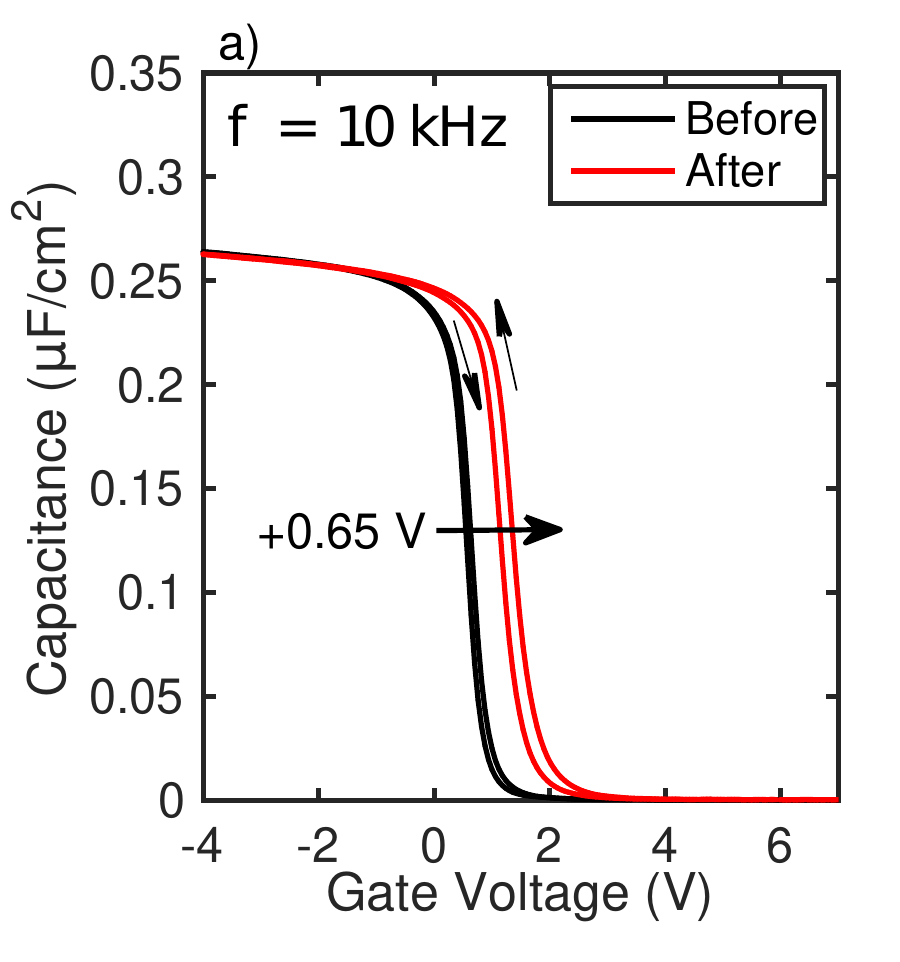}
\includegraphics[scale=0.48,trim={0.85cm 0 0cm 0cm},clip]{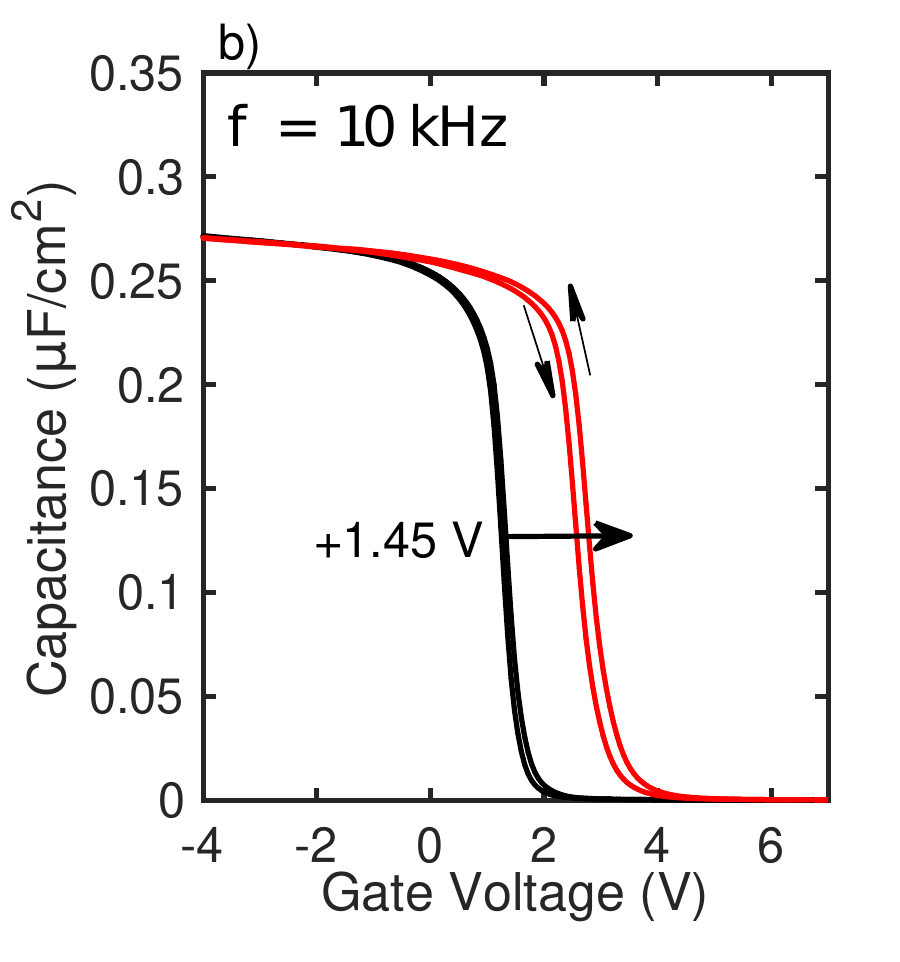}
\caption{CV measurements on the MISCAPs before and after irradiation of (a) AlOx:D1 and (b) AlOx:D2. A positive shift in the flatband voltage is observed, totaling (a) $\Delta V_{FB}\approx 0.65$~V and (b) $\Delta V_{FB}\approx 1.45$~V. This is indicative of a negative charge build-up post-irradiation. This negative charge density accumulation is estimated to be (a) 1.05$\times$10$^{12}$~cm$^{-2}$ and (b) 2.35$\times$10$^{12}$~cm$^{-2}$.}
\label{CV_prepostIrrad}
\end{figure}

The ID-VD measurements of the samples before and after irradiation are shown in Fig.~\ref{SC_al2o3_IDVD_prepostIrrad}. These results demonstrate a clear degradation in the saturation drain current post-irradiation exposure. The degradation continues with higher proton fluences. As shown in Table~\ref{table:Hall-effect_Al2O3_mu}, a reduced hole mobility is believed to explain this degradation, caused by an increase in scattering induced by band distortions along the 2D hole well. The source of these distortions arise from the Coulombic interactions with the negatively-charged states in the Al$_2$O$_3$ which increase post-irradiation. 

To further explore the negative charge build-up in the oxide, we conducted CV measurements. The results are shown in Fig.~\ref{CV_prepostIrrad}. The flatband voltage was extracted using the second derivative method (i.e., d$^2$C(V)/dV$^2\approx$0 when V equals to the flatband voltage V$_{FB}$~\cite{winter_capFB}). Post-irradiation, the flatband voltage shift $\Delta$V$_{FB}$ was 0.65 and 1.45~V for samples AlOx:D1 and D2, respectively. This is indicative of a negative charge build-up in the oxide bulk, which can be estimated using the known expression $Q = C_{ox}\Delta V/q$, where $Q$ is the charge density, $C_{ox}$ is the oxide capacitance, $\Delta V$ is the voltage shift, and $q$ is the elementary charge magnitude. This expression yields 1.05$\times$10$^{12}$ and 2.35$\times$10$^{12}$~cm$^{-2}$ for $\Delta$V$_{FB}=0.65$ and $1.45~V$, respectively. Moreover, we can gain insight on the trapping behavior at (or near) the Al$_2$O$_3$/H:diamond interface by measuring the hysteresis behavior. From Fig.~\ref{CV_prepostIrrad}, we can observe a counter-clockwise hysteresis behavior (as a $p$-type device, the sweeping was performed from positive to negative V$_G$). For both samples, the hysteresis pre-irradiation was approximately 80~mV. Post-irradiation, the hysteresis for samples AlOx:D1 and D2 increased to approximately 210 and 220~mV, respectively. Interestingly, this increase was almost identical for both doses. This may indicate that the accessible trap states at (or near) the Al$_2$O$_3$/H:diamond interface increase and saturate after a particular dose.

\begin{table*}
\centering
\caption{Hall-effect mobility and sheet density values for samples AlOx:D1 and AlOx:D2 pre/post irradiation.}
 \begin{tabular}{ | m{1.5cm} | m{2cm} |  m{2cm} | m{2cm} | m{1.5cm} |  m{2cm} | m{2cm} | m{1.5cm} | }
 \hline\hline
Substrate & Fluence (H+/cm$^{2}$) & Pre-Irrad. $\mu_H$~(cm$^{2}$/(Vs)) & Post-Irrad. $\mu_H$~(cm$^{2}$/(Vs)) & \%Change in $\mu_H$ & Pre-Irrad. n$_{2D}$~(cm$^{-2}$) & Post-Irrad. n$_{2D}$~(cm$^{-2}$) & \%Change in n$_{2D}$\\ [0.5ex] 
 \hline
AlOx:D1 & 0.5$\times$10$^{14}$ & 48.5 & 34.2 & -29.5 & 1.51$\times$10$^{12}$ &  1.81$\times$10$^{12}$ & +19.9 \\ 
AlOx:D2 & 1.0$\times$10$^{14}$ & 49.7 & 32.0  & -35.6 & 3.40$\times$10$^{12}$ & 4.36$\times$10$^{12}$  & +28.2 \\ 
 \hline\hline
\end{tabular}
\label{table:Hall-effect_Al2O3_mu}
\end{table*}

Hall-effect devices were also fabricated on samples AlOx:D1 and D2 in order to gain further insight as to how the conduction properties of the 2DHG are influenced with radiation dose. Shown in Table~\ref{table:Hall-effect_Al2O3_mu} are the mobility values for each sample. We observe a decrease in the hole mobility, changing by -29.5\% and -35.6\% for AlOx:D1 and D2, respectively. This is likely due to enhanced remote impurity scattering induced from the negative charge build-up in the oxide. The result would therefore explain the saturation current degradation observed in Fig.~\ref{SC_al2o3_IDVD_prepostIrrad}. 

With regards to the hole sheet density, we observe that it increases with proton dose. We hypothesize that this is a direct consequence of the negative charge build-up as a function of dose, just as we have observed from the CV measurements. Effectively, the 2DHG is further accumulated after irradiation, which manifests in the Hall-effect measurements as a higher sheet density. This mechanism is shown schematically in Fig.~\ref{irrad_al2o3_schematic}(c). In addition, the irradiation may have increased the concentration of O$_i$ and Al$_\mathrm{v}$ sites, which are believed to act as acceptors~\cite{kawarada2017durability}. This, in turn, would contribute to the increased 2DHG concentration. 

On the other hand, the Hall-effect measurements show a reduction in the hole mobility. However, it is difficult to determine whether its cause is induced by the damaged oxide, or from other limitations that are intrinsic to the H:diamond structure. We therefore dig deeper in our analysis by irradiating unpassivated H:diamond samples.

Similar to the passivated samples, two air-doped and unpassivated H:diamond samples (denoted as Air:D1 and Air:D2) were proton-irradiated. Samples Air:D1 and D2 were also irradiated with 2~MeV protons at 0.5 and 1$\times$10$^{14}$~cm$^{-2}$. Thus, any radiation effects induced by a damaged oxide is disregarded. However, it is important to consider that the $p$-type surface conduction is dependent on the formation and stability of C-H surface dipoles. Since the ionization rate of a 2~MeV proton in diamond is 50~eV/nm$\mathrm{\AA}$ near the surface (or about 18~eV per lattice constant), it can be inferred that the penetrating protons can deposit enough energy to dissociate the C-H bonds. If so, this would have significant implications on the general stability of H:diamond devices (both passivated or unpassivated) in environments rich with charged radiation particles.

\begin{figure}[t!]\centering
\includegraphics[scale=0.48,trim={0cm 0 0.5cm 0},clip]{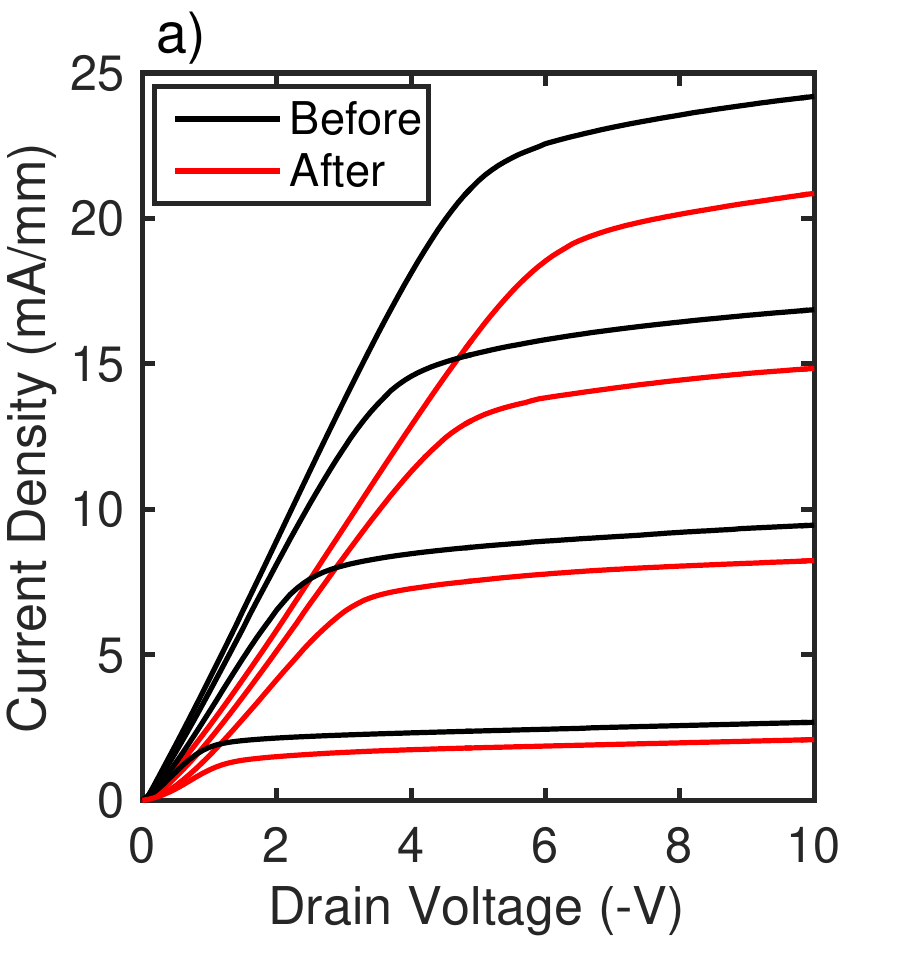}
\includegraphics[scale=0.48,trim={0.7cm 0 0cm 0},clip]{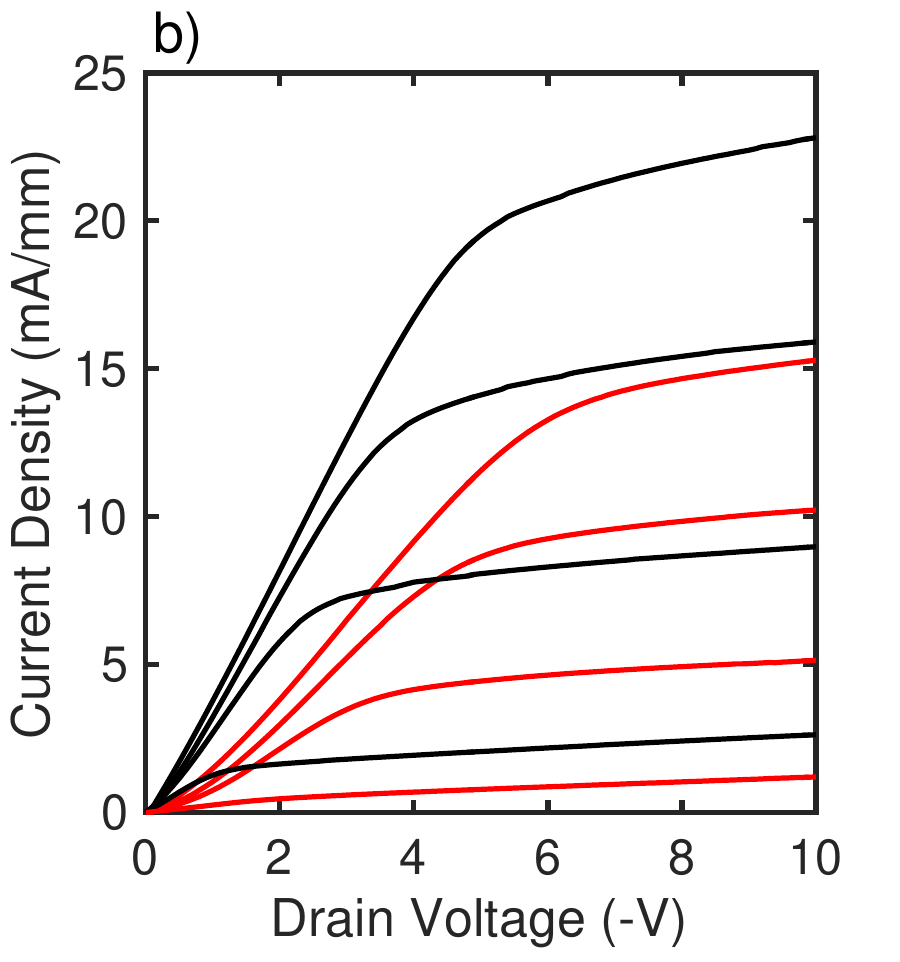}
\caption{ID-VD measurements on the MESFETs before and after irradiation of (a) Air:D1, exposed to  0.5$\times$10$^{14}$~protons/cm$^{2}$ and (b) Air:D2, exposed to 1$\times$10$^{14}$~protons/cm$^{2}$.}
\label{SC_air_IDVD_prepostIrrad}
\end{figure}

Shown in Fig.~\ref{SC_air_IDVD_prepostIrrad} are ID-VD measurements of the Air:D1 and D2 samples. We can once again observe a degradation in the drain current, which worsens with proton dose. However, this degradation rate is not as severe as the passivated samples in Fig.~\ref{SC_al2o3_IDVD_prepostIrrad}, likely owing to the absence of an enhanced scattering mechanism associated with the negative charge build-up in the oxides. However, the saturation current for the unpassivated devices also declines post-irradiation, which is suggestive of another scattering mechanism whose rate increases with proton dose.

\begin{figure}\centering
\includegraphics[scale=0.45,trim={0cm 0 .6cm 0},clip]{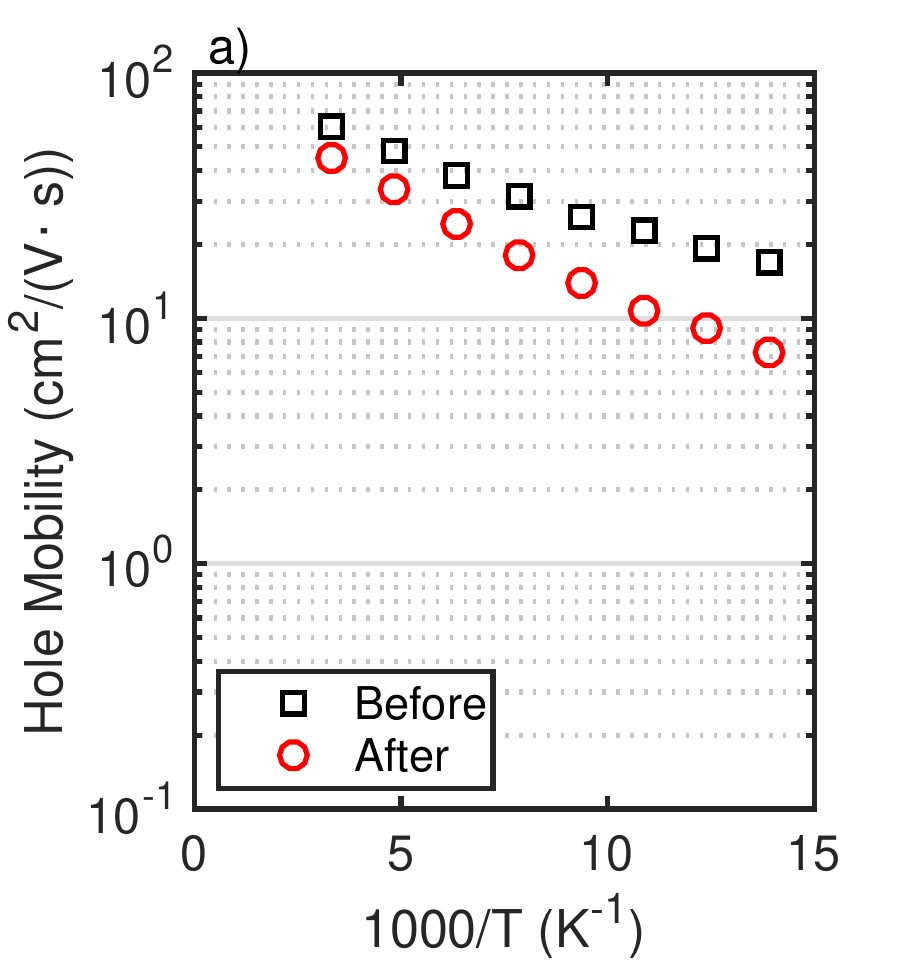}
\includegraphics[scale=0.45,trim={0cm 0 .6cm 0},clip]{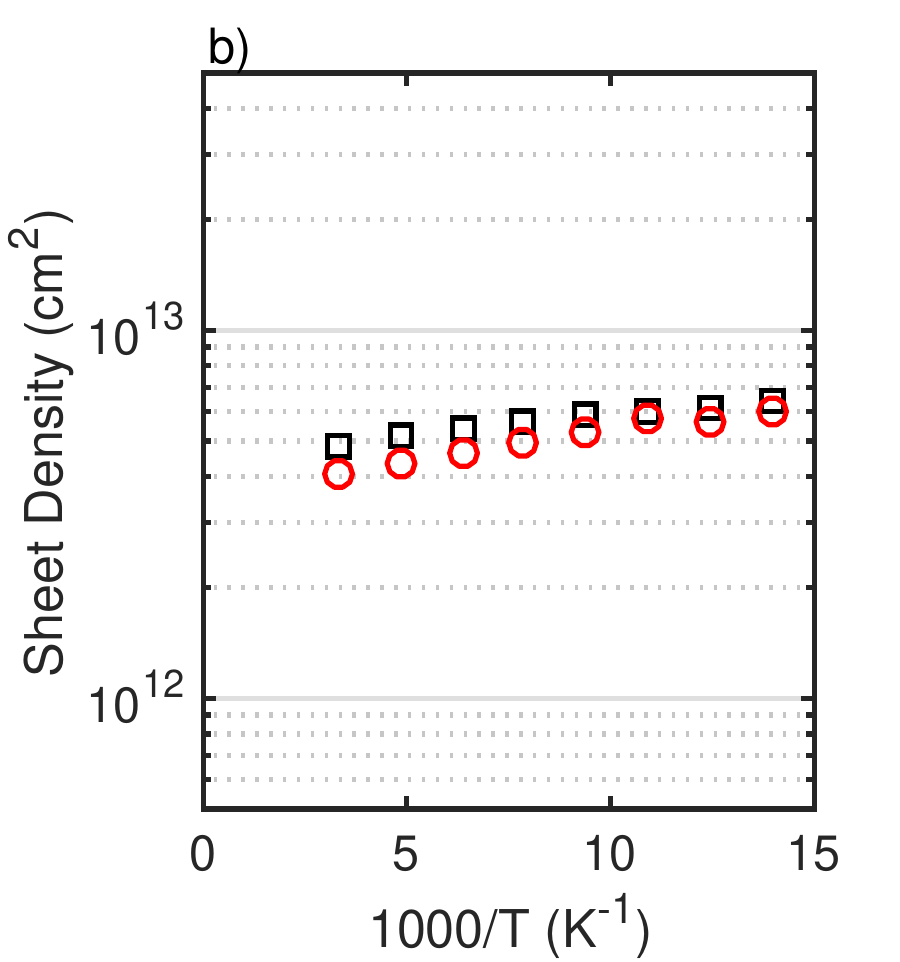}\\
\includegraphics[scale=0.45,trim={0cm 0 .6cm 0},clip]{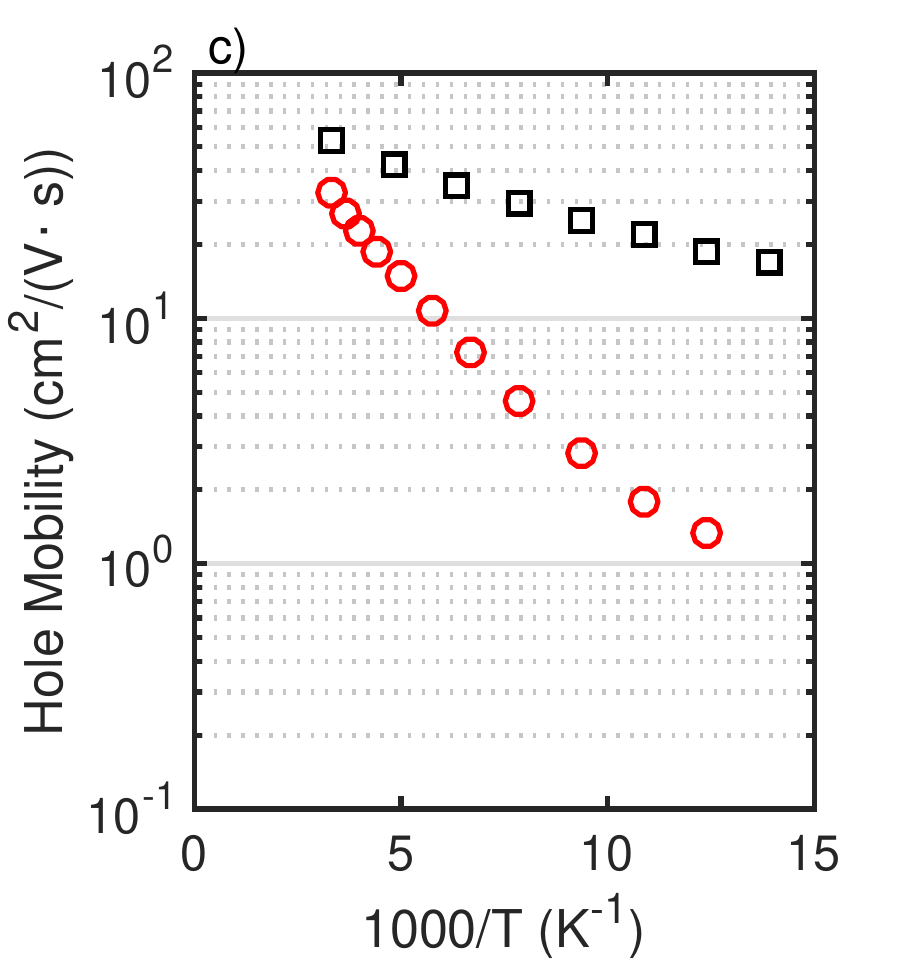}
\includegraphics[scale=0.45,trim={0cm 0 .6cm 0},clip]{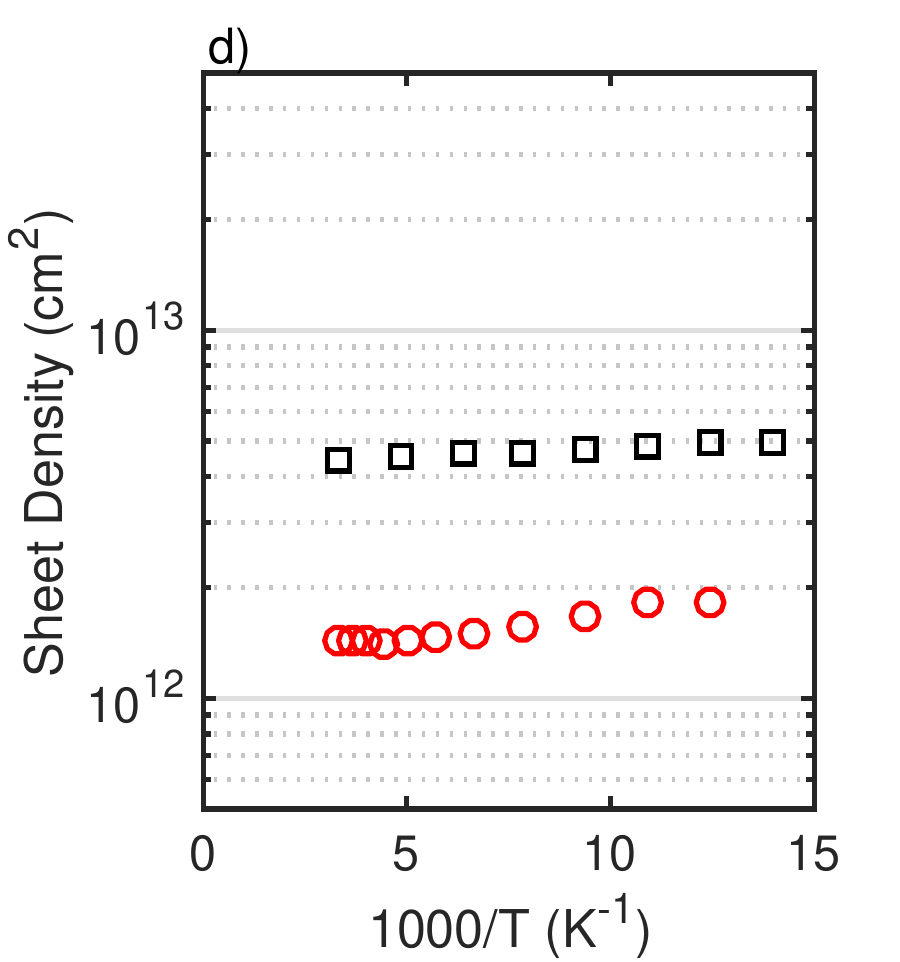}
\caption{Hall-effect measurements before and after irradiation of (a, b) sample Air:D1 and (c, d) sample Air:D2. The room temperature measurement is the first data point on each plot.}
\label{SC16_SC22_mu_n2D}
\end{figure}

To gain insight on the exact mechanisms degrading these properties, we turn to Hall-effect measurements to extract the hole mobility and sheet density from RT to 70~K. The measurements were swept bidirectionally from RT to 70~K, and negligible hysteresis was observed. This temperature dependence produced an activation behavior that provides insight into the mechanisms responsible for this degradation, which was the main thesis of our previous work \cite{Peterson2020}.

Shown in Fig.~\ref{SC16_SC22_mu_n2D}(a, c) are the Hall-effect mobilities for samples Air:D1 and D2. Here we see a notable degradation of the RT mobility for both samples, from 60.5 to 45.4~cm$^2$/(V$\cdot$s) for Air:D1, and from 53.2 to~{32.7~cm$^2$/(V$\cdot$s)}~for Air:D2. There is thus a scattering mechanism that is enhanced post-irradiation. At least two plausible mechanisms can explain this observation. First, the ionization of the penetrating protons may impart enough energy to create vacancies and interstitials throughout the diamond surface. Second, the ionization may impart enough energy to dissociate the C-H dipoles \cite{Peterson2020, crawford2021surface}. Both of these degradation mechanisms would create band distortions which would, in turn, scatter the conducting holes and reduce their mobility. 

We believe that the first of these explanations (creation of lattice vacancies and interstitials) is unlikely, given the electronic and nuclear stopping power functions for a 2~MeV proton in diamond. Using SRIM, we can simulate both the ionization rate, as well as the vacancy production rate (in vacancies/nm), for any charged ion and target material. Near the surface, SRIM estimates that each 2~MeV proton produces approximately 1$\times 10^{-4}$ vacancies/nm. That is, a proton must penetrate through 10 micrometers before a single vacancy is produced. Fortunately, an experimental study by Manfredotti~$\textit{et. al.}$ supports this reasoning~\cite{H3_2MeVprotons_SCD}. These authors also exposed 2~MeV protons to single crystal diamond samples (which were purchased from the same manufacturer that supplied the samples for this study). During and after proton exposure, Manfredotti~$\textit{et. al.}$ performed luminescence measurements and found that the peak for the H3 center (regarded as the center for vacancies and interstitials in diamond, and is commonly enhanced by irradiation~\cite{H3center1,H3center2,H3center3}) increases marginally after a fluence of 9$\times 10^{14}$~cm$^{-2}$. This is 9-fold the maximum fluence in this study. Moreover, these luminescence techniques collect light from a depth of multiple microns. Since the induced lattice damage increases with the depth of the ion, this means that the signal for the H3 center would be even weaker near the superficial regions where the 2DHG exists. For this reason, it is doubtful that the lattice defects are the primary reason behind the degraded mobility measured in samples Air:D1 and D2.

However, ionization of the surface may impart sufficient energy to dissociate the C-H bonds, which we know are necessary for the formation of the 2DHG. Other studies on the irradiation of H:diamond with neutrons~\cite{verona2016_neutrons}, as well as gamma~\cite{gamma_on_Hdiamond}, show no evidence of degradation. This is likely because these uncharged radiation species do not emit a trail of ionization energy as charged particles do. There have been studies of keV-electrons dissociating the C-H bonds, however~\cite{cui1999electron}. Moreover, the charged particle radiation effects on surfaces and interfaces have been studied extensively for a wide variety of materials. For example, it is well established that the adhesion of metal/ceramic interfaces increase significantly after irradiating with MeV ions~\cite{livi1985_metaladhesion,werner1983_metaladhesion}, which is attributed to the interactions caused by ionization. C-H bonds have also been shown to dissociate upon MeV irradiation in a multitude of hydrocarbon-rich materials ~\cite{tombrello1987ubiquity}. In general, in multiple contexts involving MeV-ion irradiation, the ionization-induced desorption of surface molecules is widely inferred to explain various phenomena~\cite{surface_mod_MeV_ions,Interaction_w_surfaces}. 

\begin{figure}[t!]\centering
\includegraphics[scale=0.35,trim={0cm 0 0cm 0},clip]{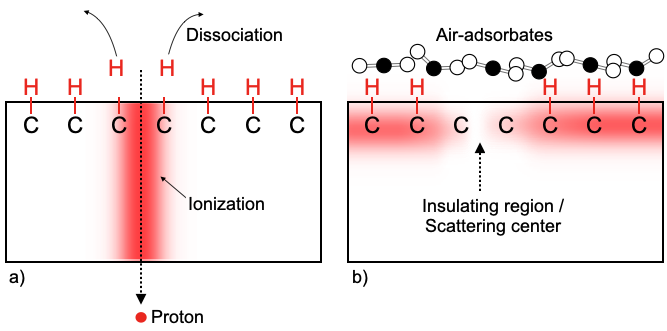}
\caption{Schematic demonstrating the dissociation of C-H bonds resulting from proton ionization of (a) Proton irradiation of H:diamond in the vacuum chamber and (b) the induced 2DHG after air exposure. Any dissociation of the C-H surface would degrade the 2DHG electrical properties.}
\label{ch_diss_radiation}
\end{figure}

As we discussed in our previous work \cite{Peterson2020}, as well as the work by Garrido~$\textit{et. al}$~\cite{Garrido_2005}, it is believed that long-range potential fluctuations act as scattering centers in H:diamond, which may in turn arise from incomplete hydrogen termination. Along this line of reasoning, should the ionization induced by the protons dissociate the C-H bonds, more scattering centers would be produced along the 2D hole well, which would explain the degraded mobility and saturation current. This is shown schematically in Fig.~\ref{ch_diss_radiation}.

With respect to the irradiation effects on the 2DHG sheet density, this is shown in Fig.~\ref{SC16_SC22_mu_n2D}(b, d). We see from both samples that the sheet density degrades. For the sample Air:D1, we observe that the degradation is small, yet noticeable, from 4.86 to 4.06~$\times10^{12}$~cm$^{-2}$, which is a 16.5\% decrease. On the other hand, the sheet density degradation for Air:D2 is much more, from 4.46 to 1.45~$\times10^{12}$~cm$^{-2}$, which is a 67.0\% decrease. The general trend here is that the 2DHG density degrades with fluence, which is presumably caused by the dissociated C-H bonds. It is worth noting that the large drop in sheet density between samples Air:D1 and D2 is unexpected, given that the latter dose is only 2x greater. Since these are separate samples, there may be intricacies of the surface that explain such a large drop. Alternatively, there may be a degradation mechanism with a threshold between both fluences, which could only be interpolated with more fluence data points between 0.5 and 1$\times 10^{14}~\mathrm{cm}^2$. Therefore, more detailed studies of charged particle radiation effects on H:diamond are recommended. 

In conclusion, we discussed the work of irradiating four single-crystal diamond samples, two of which were passivated with Al$_2$O$_3$, and the other two left unpassivated. Several devices were fabricated and measured prior to irradiation, which include MISFETs, MISCAPs, MESFETs and Hall-effect devices. The samples were irradiated with 2~MeV protons at two fluences, 0.5 and 1$\times$10$^{14}$~cm$^{-2}$. In order to decouple the degradation that is intrinsic to the H:diamond from the degradation of the passivation layer, the characteristics of MESFETs and Hall-effect devices of unpassivated H:diamond samples were exposed and measured. 

It was determined that the Al$_2$O$_3$ films were negatively charged post-irradiation, which consequently degraded the hole mobility, but also increased the sheet density via electrostatic attraction. The net result was a substantial decrease in the drain saturation currents, and shifts in the threshold and flatband voltages. It was also determined that the hole mobility and sheet density drop on the unpassivated H:diamond samples. The net result was also a decrease in the drain saturation current. Since MeV-protons exhibit negligible nuclear scattering near the surface, and deposit their energy entirely via ionization, we suspect that the degraded hole mobility is caused via enhanced scattering off insulating barrier centers throughout the 2D hole well.

See supplementary material for a further discussion on how the temperature dependence of the Hall-effect measurements reveals how the H:diamond surface and the 2DHG is affected by proton irradiation. 

This work was supported by the National Science Foundation (NSF) Graduate Research Fellowship under Grant No. DGE-1656518. Part of this work was performed at the Stanford Nanofabrication Facility (SNF) and the Stanford Nano Shared Facilities (SNSF), which is supported by NSF under Award No. ECCS-1542152. Many thanks to the SNF staff for their support.

\nocite{*}
\bibliography{aipsamp}

\end{document}